\documentclass[a4paper,fleqn,usenatbib]{mnras}

\usepackage[T1]{fontenc}
\usepackage{ae,aecompl}

\usepackage{graphicx}	
\usepackage{amsmath}	
\usepackage{amssymb}	

\def\noao{The NOAO Deep Wide Field Survey MOSAIC Data Reductions, 
http://www.noao.edu/noao/noaodeep/ReductionOpt/frames.html}%
\title[Radial profile of the LMC]{Stellar density distribution along the minor axis of 
the Large Magellanic Cloud}

\author[Piatti et al.]{
Andr\'es E. Piatti$^{1,2}$\thanks{E-mail: andres@oac.unc.edu.ar} 
\\
$^{1}$Consejo Nacional de Investigaciones Cient\'{\i}ficas y T\'ecnicas, Av. Rivadavia 1917, 
C1033AAJ, Buenos Aires, Argentina\\
$^{2}$Observatorio Astron\'omico, Universidad Nacional de C\'ordoba, Laprida 854, 5000, 
C\'ordoba, Argentina\\
}

\date{Accepted XXX. Received YYY; in original form ZZZ}

\pubyear{2017}

\begin{document}
\label{firstpage}
\pagerange{\pageref{firstpage}--\pageref{lastpage}}
\maketitle

\begin{abstract}
We studied the spatial distribution of young and old stellar populations along the western 
half part of the minor axis of the Large Magellanic Cloud (LMC) using Washington $MT_1$
photometry of selected fields, which span a deprojected distance range from the LMC bar centre
out to $\sim$ 31.6 kpc. We found that both stellar populations share a mean LMC limiting 
radius of 8.9$\pm$0.4 kpc; old populations are three times more dense that young populations
at that LMC limit. When comparing this result with recent values for the 
LMC extension due to north, the old populations resulted significantly more elongated than the 
young ones. Bearing in mind previous claims that the elongation of the outermost LMC regions
may be due to the tidal effects of the Milky Way (MW), our  findings suggest that such a tidal
interaction should not have taken place recently. The existence of young populations
in the outermost western regions also supports previous results about ram pressure stripping 
effects of the LMC gaseous disc due to the motion of the LMC in the MW halo.
\end{abstract}




\section{Introduction}

The advent of wide-field imaging databases has recently exploited to provide 
new insights about the geometry and extension of the Large Magellanic Cloud (LMC)
\citep[e.g.][]{beslaetal2016,jacyszynetal2017}. The direction due to north
from its centre has been explored \citep[][]{betal15,mackeyetal2016}, as well as
that towards the Magellanic Bridge (MB) \citep[e.g.][]{setal14,wagnerkaiser17}. 
The main results reveal a smooth transition of the metallicity and distance 
distributions of MB RR Lyrae stars that connect both Magellanic Clouds, and 
an intrinsic eccentricity of the old field star distribution in the LMC outer
regions with its major axis  roughly pointing towards the Milky Way (MW). 

The western half part of the minor axis of the LMC is at an intermediate direction between 
those to the north and 
towards the MB. As far as we are aware, it has not been considered  for estimating the 
LMC extension. Nevertheless,   it could shed new light in the context of the tidal interactions of the 
LMC with the MW and the Small Magellanic Cloud (SMC). 
For instance, it is opposite to the direction of  the motion of the LMC in the MW halo 
\citep{beslaetal07,kallivayaliletal13}; the H\,I column density map from
the Parkes Galactic All-Sky Survey \citep[GASS,][]{kh2015} shows extended regions of gas 
there, in contrast 
with its distribution in the northern outermost LMC disc; the
isodensity curves of field stars \citep{vdmarel2001} and 
star clusters \citep{betal08,siteketal2016} are seemingly compressed towards this
direction respect to 
the eastern side of the galaxy, among others.

In this work, we have taken advantage of the Washington  photometric data set 
described in \citet{petal2017} and used elsewhere  \citep{pc2017,p17b}. 
From Hess diagrams (Section 2), we constructed the deprojected stellar 
density distributions for young and old field stars (Section 3). We  analysed 
in Section 4 their
differences and compared our results with stellar density profiles derived
from the Dark Energy Survey (DES) data sets. Finally, in Section 5, we summarize the main
conclusions of this work.

\section{Overview of the data set}

\begin{figure}
\includegraphics[width=\columnwidth]{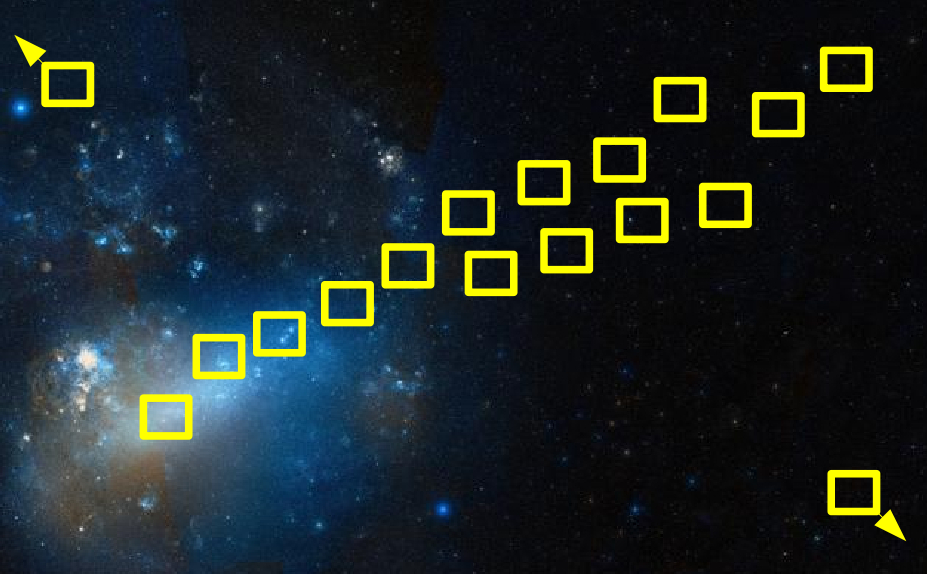}
    \caption{Spatial distribution of the presently studied LMC star fields overplotted on a image 
downloaded from the DSS2 All Sky Survey via Wikisky (http://www.wikisky.org/).
    }
   \label{fig:fig1}
\end{figure}

The whole data set comprises 17 different LMC fields 
(36$\arcmin$$\times$36$\arcmin$ wide each) distributed along the LMC minor axis
as illustrated in Fig.~\ref{fig:fig1}.  They were imaged through the  
Washington $CM$ and Kron-Cousins $R$ filters with the Mosaic\,II camera attached at 
the Cerro-Tololo Inter-American Observatory (CTIO) 4 m Blanco telescope 
during the nights Dec 27-30, 2008 
(typical seeing $\sim$ 1.1$\arcsec$, airmass range $\sim$ 1.2-1.6). The 
Kron-Cousins $R$ filter was employed as the recommended substitute of the Washington
$T_1$ filter \citep{g96}.

The data processing included different tasks, among them, to perform overscan, trimming, bias subtraction and flat-field corrections and to update the world coordinate system (WCS),
following the procedures documented by the NOAO Deep Wide 
Field Survey team \citep{jetal03}. In order to standardize the stellar
photometry we measured nearly 270 independent 
magnitudes in the standard fields PG0321+051, SA\,98 and SA\,101 \citep{l92,g96}, observed
three times per night. We derived mean extinction and colour term coefficients for the four nights
of (0.29,-0.09), (0.14,-0.24) and (0.08,-0.02) and rms errors of 0.030, 0.021, and 0.024 
in $C$, $M$, and $T_1$, respectively.

The stellar photometry for each single mosaic was obtained
after deriving the respective quadratically varying point-spread-function (PSF). 
The procedure to derive PSF magnitudes of the stars identified in each field
consisted in applying the resultant PSF and identifying
new fainter stars in the output subtracted frame to enlarge the star sample. 
We iterated this loop three times. The photometric errors were estimated
from artificial star tests performed on synthetic images created by adding nearly
5 per cent of the measured stars. For the sake of the readers we refer to \citet{petal2017}.

\section{LMC star field Hess diagrams}
 
We used the Washington $T_1$ versus $M - T_1$ colour-magnitude diagrams (CMD) to delineate the
regions where to count the number of stars per field, and thus to produce the
stellar density distribution as a function of the deprojected distance  from the LMC centre. 
Because of the number of stars measured per magnitude interval depends on the
completeness of the photometry, we first computed the respective completeness fractions. 
To do this, we used the stand-alone {\sc addstar} program in the {\sc daophot}
package \citep{setal90} to add synthetic stars,
generated at random with respect to position and magnitude, to the deepest images in order 
to derive their completeness levels. We added a number
of stars equivalent to $\sim$ 5$\%$ of the measured stars in order to avoid in the synthetic images 
significantly more crowding than in the original ones. 
On the other hand, to avoid small number statistics in the artificial-star 
analysis, we created a thousand different images for each original image. We used the option 
of entering the number of  photons per ADU in order to properly add the Poisson noise to the star 
images. 

\begin{figure}
\includegraphics[width=\columnwidth]{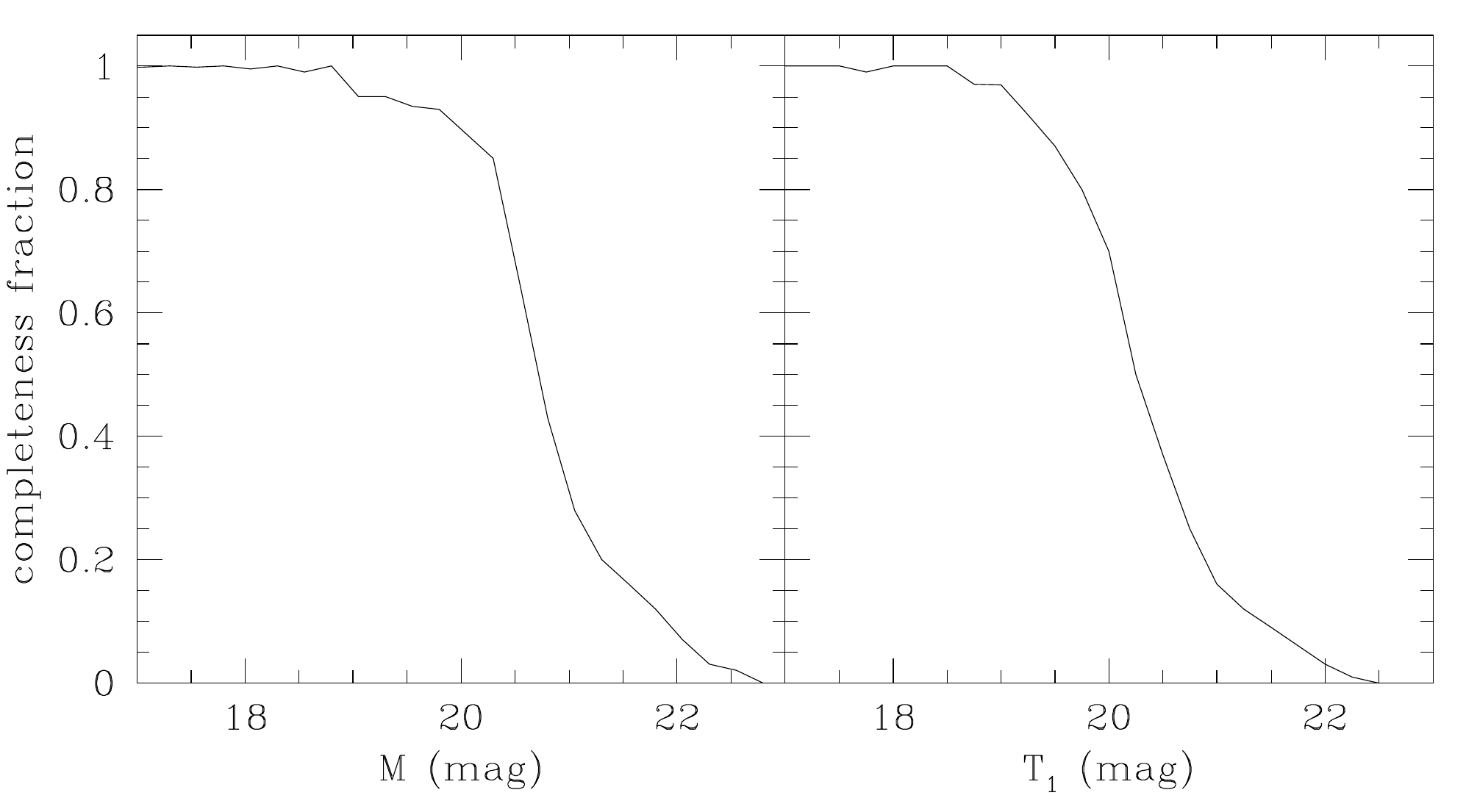}
    \caption{Completeness fraction per filter for the
crowdest field in our sample.}
   \label{fig:fig2}
\end{figure}

We then performed stellar photometry using the star-finding and point-spread-function (PSF) fitting 
routines in the {\sc daophot/allstar} suite of programs \citep{setal90}.
 We then used the {\sc allstar} program to apply the previously obtained PSF to the 
identified stellar objects and to create a subtracted image which was used to find and measure
 magnitudes of additional fainter stars. This procedure was repeated three times per image. 
The star-finding efficiency was estimated by comparing the output 
and the input data for these stars using the {\sc daomatch} and {\sc daomaster} tasks.
We illustrate in Fig.~\ref{fig:fig2} the resultant completeness fraction for the densest studied
field, located close to the midst of the LMC bar.

The global properties of the CMDs are depicted in Fig.~\ref{fig:fig3} as Hess diagrams
(the darker a CMD region the more numerous the stellar population at that position), so that
the frequency or density of occurrence of stars at various positions can also be seen.
The figure also includes the errorbars as recovered from the photometry of synthetic stars.
The Hess diagrams present the signature of the MW foreground
field which blurs particularly the LMC outskirts main features. Background galaxies have been excluded
by constraining the  photometric data sets to objects with $\chi$ $<$ 2, photometric error less than 
2$\sigma$ above the mean error at a given magnitude, and $|$SHARP$|$ $<$ 0.5.
Old main sequence (MS) populations are seen in the outermost fields, and increasing younger populations 
(brighter $T_1$ MS turnoffs) for inner LMC fields. Thus, the series of panels depicted in 
Fig.~\ref{fig:fig3} show the age transition of the stellar populations along the western side of
the minor axis of the LMC.

In order to clean the CMDs from the MW contamination we generated synthetic CMDs for each LMC field
using the Besan\c{c}on galactic model \citep{retal03}, thus coping with the MW star
field variation with the Galactic latitude. The ingredients we took into account are
as follows: an area equal to that of our studied fields and the respective Galactic coordinates, 
a distance interval from 0 up to 50 kpc in steps of 0.5 kpc, a recommended mean diffuse absorption of 
0.7 mag/kpc, no restriction in the absolute magnitude, MK spectral types and  luminosity, and the 
Johnson-Cousins $V$ and $R$  ($\equiv$ $T_1$) filters. 
We scaled the synthetic CMDs using as reference
the number of stars per $T_1$ magnitude and $M-T_1$ colour intervals counted in the CMDs of
two fields located well beyond the LMC (see boxes  with an arrow in Fig.~\ref{fig:fig1}).
Because of the position of these two fields, they were suitable  for checking variations in the
MW CMDs. 

\begin{figure*}
\includegraphics[width=\textwidth]{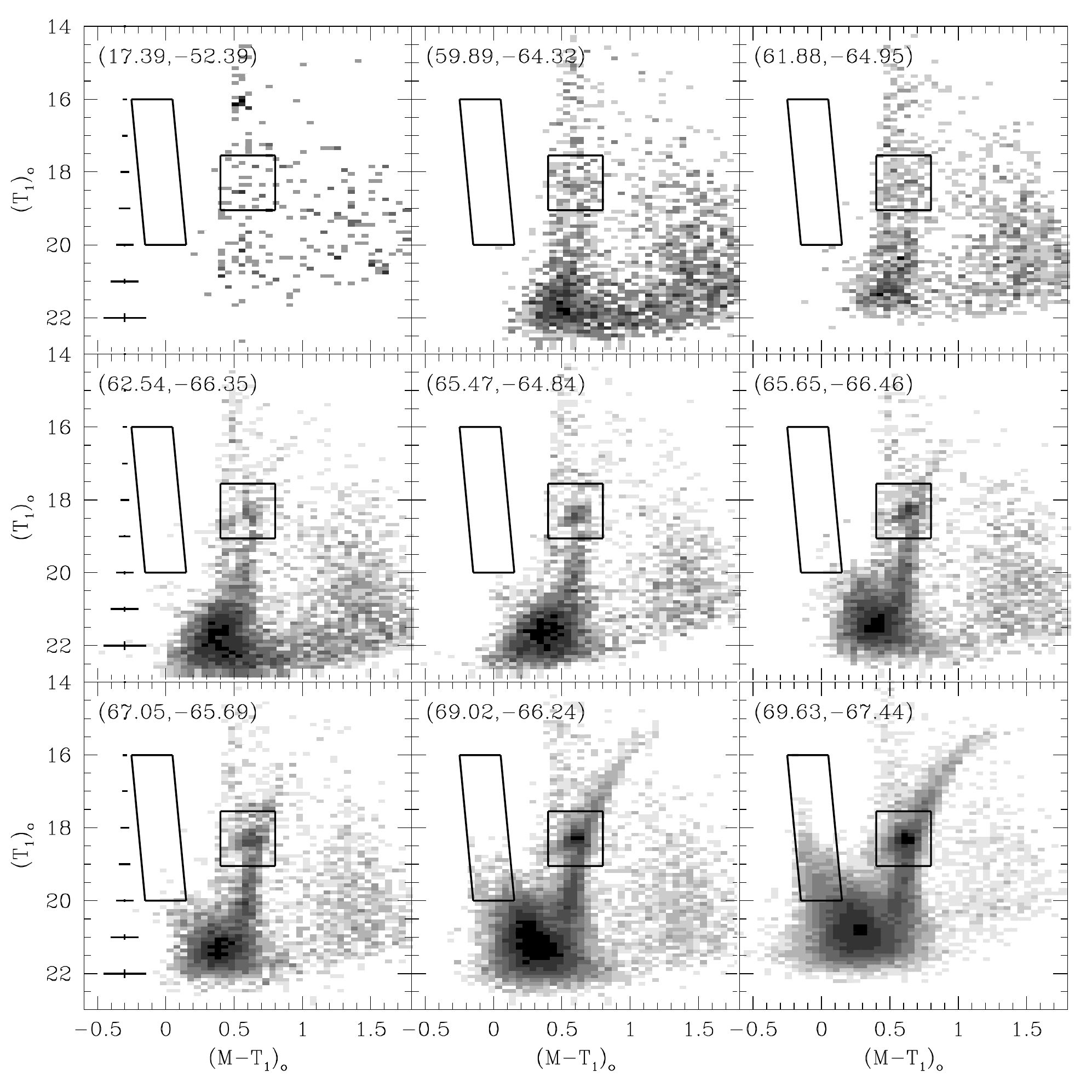}
    \caption{Dereddened $T_1$ vs. $M-T_1$ Hess diagrams of the studied fields. Central RA, Dec. coordinates 
(degrees) are indicated at the top-left margin of each panel. Errorbars are indicated at the left
margin. The defined regions along the MS and at the RC are also superimposed.}
   \label{fig:fig3}
\end{figure*}

\setcounter{figure}{2}
\begin{figure*}
\includegraphics[width=\textwidth]{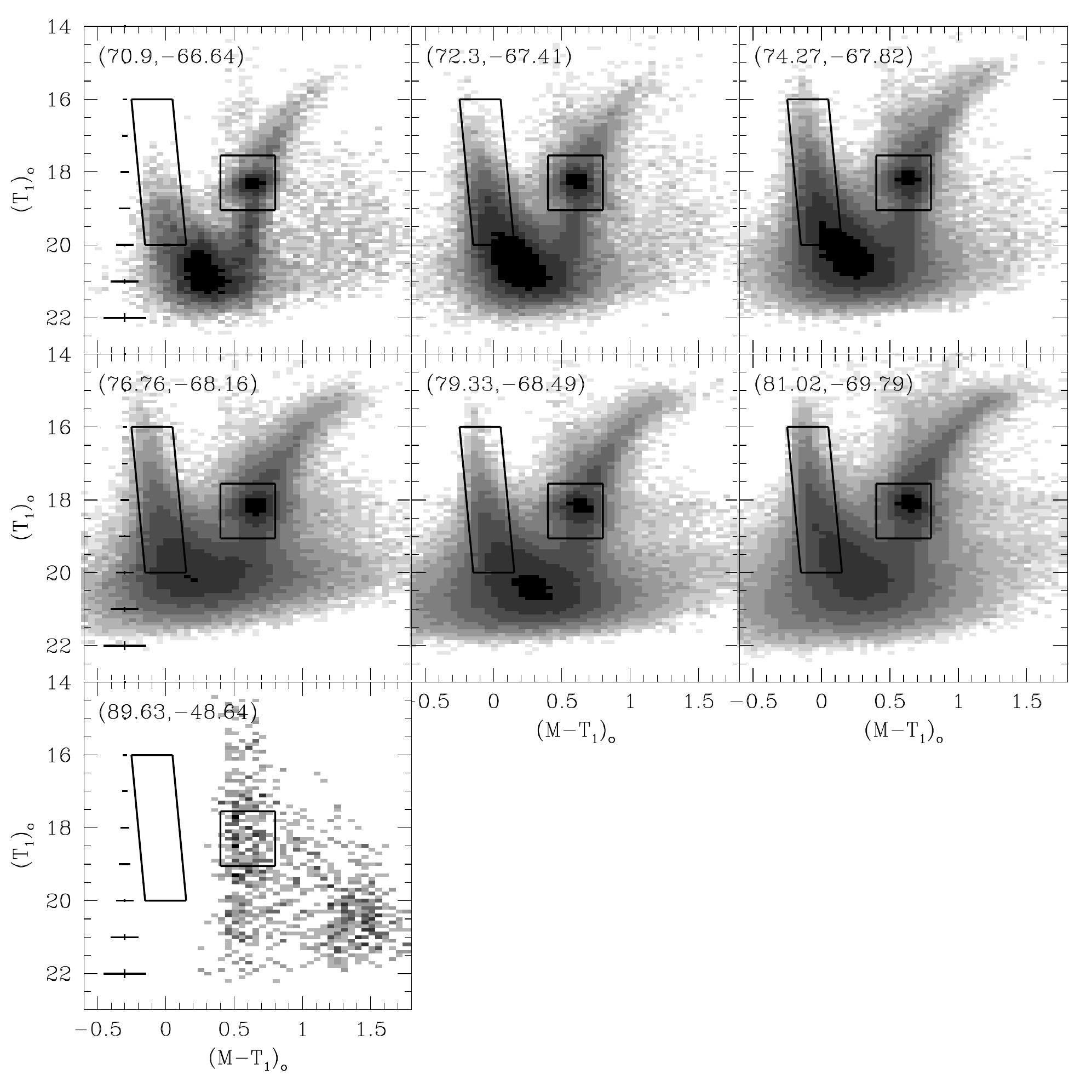}
    \caption{continued.}
\end{figure*}

\section{Stellar density profiles}

We counted the number of stars distributed along the MS and at the red clump (RC) as
representatives of the stellar densities of young (age $<$ 1 Gyr) and old (age $\ge$ 1 Gyr) 
populations \citep[see,][]{betal15}. The MS strip is delimited by the straight line $(T_1)_o$ = 
40.0$\times$$(M-T_1)_o$ + 26.0 and a parallel line shifted $\Delta$($(M-T_1)_o$) = 0.3 mag towards
redder colours, and the $(T_1)_o$ magnitudes 16.0 and 20.0, respectively. As for the RC, we chose a box 
delimited by $(M-T_1)_o$ = 0.40 and 0.80 mag, and $(T_1)_o$ = 17.55 and 19.05 mag, respectively.
Note that by using the $M-T_1$ colour instead of $C-M$ or $C-T_1$ -also available in our 
database-, reddening and metallicity effects are minimized \citep{c76}. Fig.~\ref{fig:fig3}
depicts these defined CMDs regions for comparison purposes.

These two CMD regions were selected on the basis of the positions of theoretical isochrones of
\citet{betal12} with ages and metallicities embracing the values known for the age-metallicity
relationship of the LMC \citep{pg13}, namely: for young populations we considered [Fe/H] values
between -0.5 and -0.1 dex, while for the old populations we expanded the range -1.2 - -0.1 dex.

Since we dealt with CMDs affected by different amounts of interstellar extinction, we interpolated $E(B-V)$ values
in the Magellanic Clouds (MCs) extinction map obtained from the RC and  RR Lyrae stellar photometry 
provided by the OGLE\,III collaboration \citep{hetal11}, and used the $E(B-V)$ 
values provided by the NASA/IPAC Extragalactic Database\footnote{http://ned.ipac.caltech.edu/. 
NED is operated by the Jet Propulsion 
Laboratory, California Institute of Technology, under contract with NASA.} (NED) for fields located
beyond the area covered by the MCs extinction map. From them we properly
shifted the adopted MS strip and RC box prior counting the number of stars lying within them.
We recall that \citet{hetal11} found very low reddenings in the LMC bar region \citep[see, also,][]{p17b}.
In performed this task we used the equations 
$E(M-T_1)$/$E(B-V)$ = 0.90 and $A_{T_1}$/$E(B-V)$ = 2.62 \citep{g96}. 
As can be seen in Fig.~\ref{fig:fig3}, with the adopted $E(B-V)$ values, the defined MS strip and 
RC box  encompass well the respective observed stellar population.

We also corrected by  incompleteness effects the number of stars counted, using the respective completeness
curves (see Fig.~\ref{fig:fig2}). Note that we used stars with magnitudes much brighter than that for 
the 50 per cent completeness level.
We followed the same procedure described above for counting the number of stars located within the selected regions in the MW synthetic CMDs, which we subtracted from the observed LMC stellar
density  profiles. 

Uncertainties in the star counts come from the fact that a star, because of its photometric errors,
can fall outside the defined MS strip or RC box, or vice-versa, so that the total amount of stars
can change.  Therefore,  we considered the photometric errors $\sigma(T_1)$ and $\sigma(C-T_1)$ 
of each star  inside the MS strips and RC boxes to estimate the errors in the respective star
counts. In the case of the MW subtracted LMC stellar density  profiles we considered
in quadrature both the errors in the MW profile and those in the observed LMC  ones.

\section{analysis}

\begin{figure*}
\includegraphics[width=\columnwidth]{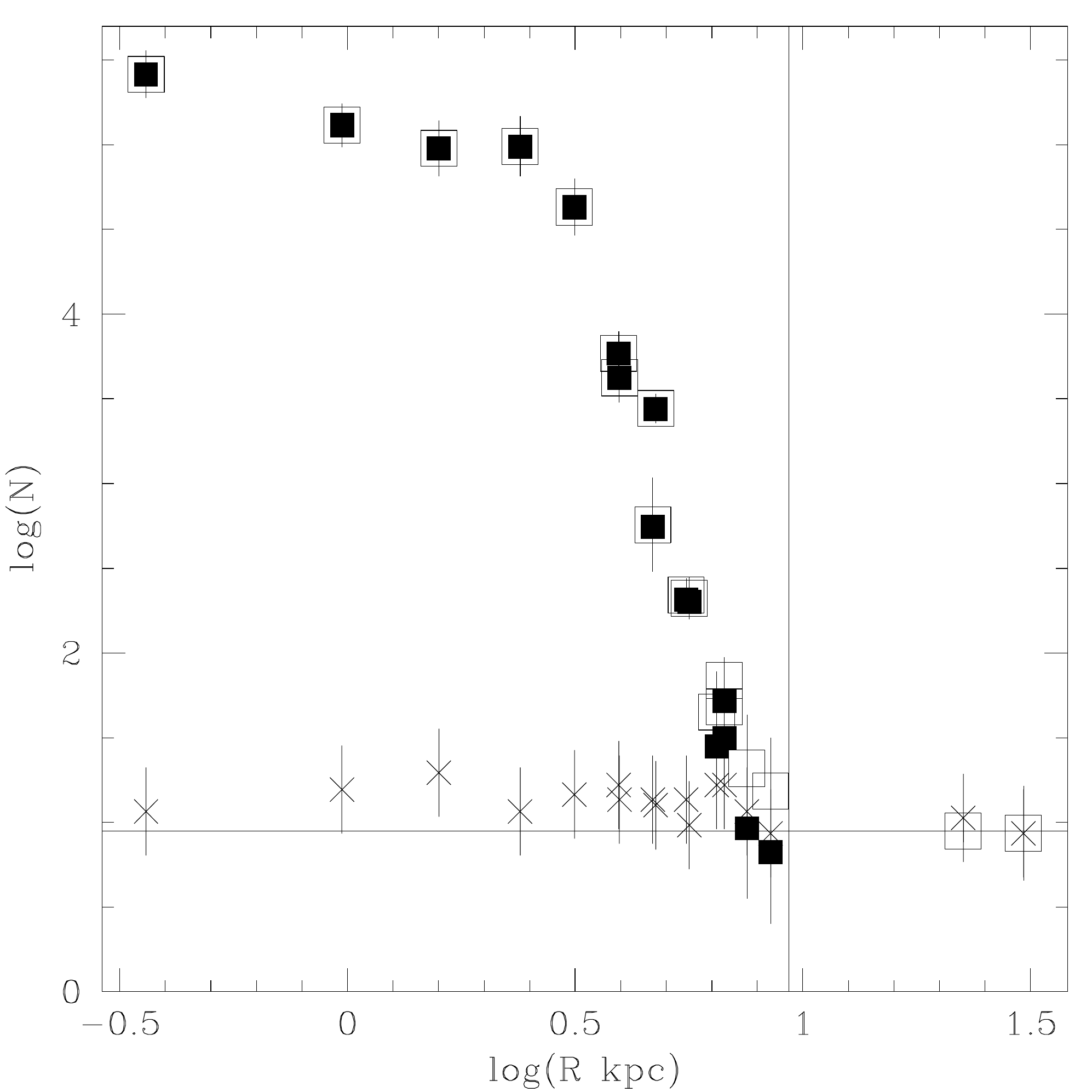}
\includegraphics[width=\columnwidth]{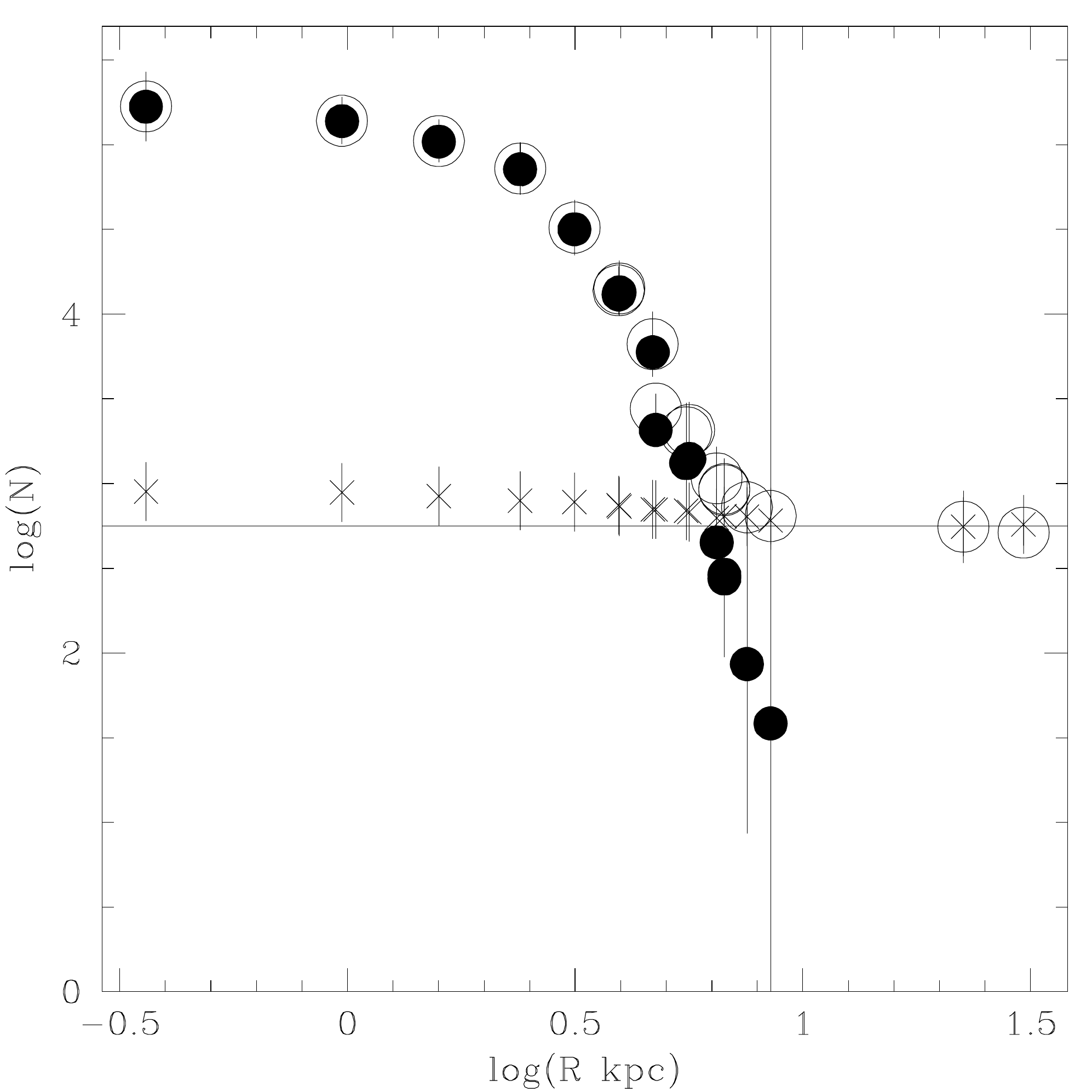}
\includegraphics[width=\columnwidth]{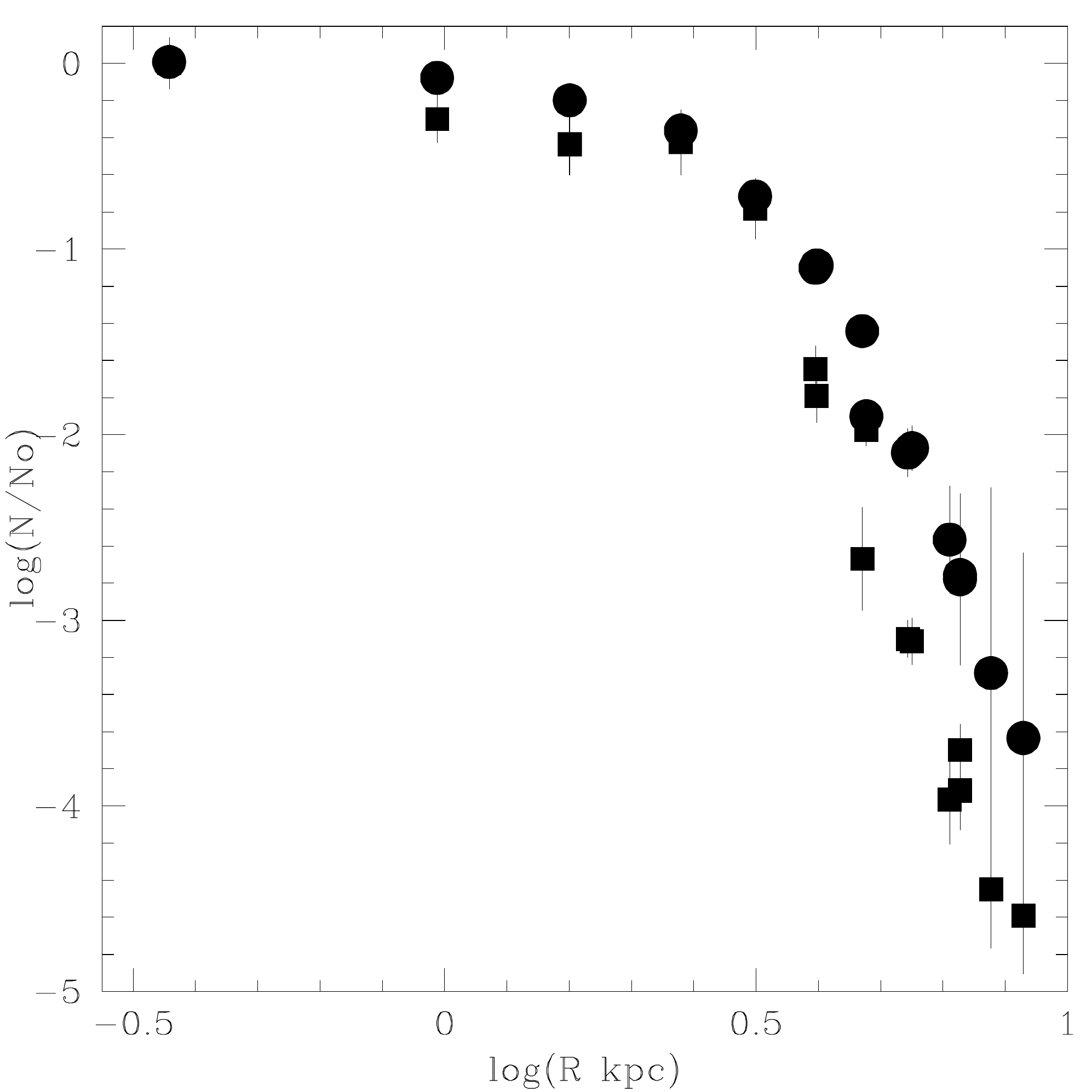}
    \caption{Observed (open symbols) and MW decontaminated (filled symbols) stellar density 
    profiles for young (left panel) and old (right panel) stellar populations as a
    function of deprojected distances.
        Crosses represents those for the MW. The mean MW
    stellar densities beyond the periphery of the LMC (horizontal lines) and their 
    respective intersections  with the observed LMC profiles (vertical lines) are overplotted.
    The bottom panel shows the normalized MW decontaminated stellar density profiles for
    comparison purposes.
    }
       \label{fig:fig4}
\end{figure*}

Fig.~\ref{fig:fig4} depicts the resultant stellar density profiles for young and old populations
as a function of the deprojected distance. The latter were computed assuming an LMC disc
with an inclination of 38.14$\degr$, a position angle of the line-of-nodes of 129.51$\degr$
and centred at R.A. = 05$^{\rm h}$ 23$^{\rm m}$
  34$^{\rm s}$, Dec. = $-$69$\degr$ 45$\arcmin$ 22$\arcsec$ (J2000) \citep{betal15}.
  For the sake of the reader we also included those for the MW, which we subtracted from the
observed LMC profiles to obtain the MW decontaminated ones. As can be seen, the stellar
density variation of the MW is not significant across the observed LMC sky, and 
the subtraction of it becomes more important in the outskirts of the LMC, where the number of 
MW MS and RC stars are of the same order than those counted in the most distant fields from the 
LMC centre. This effect is taken into account by the size of the errorbars.

The most striking feature shown by Fig~\ref{fig:fig4} is that both young and old populations
extend similarly, although reaching different outermost stellar densities.
Both profiles result indistinguishable from the respective MW levels at a deprojected 
distance of 8.9 $\pm$ 0.4 kpc. This is not a usual picture seen towards other
directions from the LMC centre, nor in the MW and other galaxies, where old populations usually 
reach larger distances than the young ones. For instance, \citet{betal15}  using DES 
photometry along a fringe due to north found deprojected
distances of 10.0 and 15.1 kpc for young and old populations, respectively. The same data set was exploited by \citet{mackeyetal2016} 
who found that at $\sim$ 18.5 kpc there is no evidence for LMC  populations in any direction.
We checked whether the different LMC centres adopted by Balbinot et al. and Mackey et al.
and in this work could  imply a non-negligible difference in the calculated deprojected
distances. We found that neither the different centres nor the different most common values for the 
LMC inclination and position angle of the line-of-nodes result in  so different deprojected distances  as to dodge comparing them consistently.

We computed the ratio of radii towards the northern (Balbinot et al's results) and western 
minor axis directions and found values of 1.1 and 1.7 for young and old populations, 
respectively. In case of using the Mackey et al's results for the old  populations,  we
found a ratio of 2.1. These results agree well with the Luminance filter images shown by
\citet{beslaetal2016}, where the LMC seems somehow compressed
towards the western minor axis direction as compared to its northern extension, 
as well as with the isodensity curves traced from 2MASS data \citep{vdmarel2001}.

Some further spatial tracers are also interesting to consider. For instance, the 
H\,I column density map built from the GASS survey
shows that gas is not as dense concentrated towards the north as along the western minor axis, in
agreement with the relative spatial distribution of 567 OB star candidates listed by 
\citet{cdetal2012} \citep[][see their figure 7]{mbetal2017}. According to \citet{salemetal2015} 
and \citet{is2015} the H\,I spatial and velocity distributions witness the perturbation by
ram pressure effects due to the motion of the LMC in the MW halo, i.e., some amount of gas
 could have traveled in the opposite direction to that motion. Particularly, \citet{salemetal2015}
found evidence that the LMC's gaseous disc has recently experienced ram pressure stripping, with
a truncated gas profile along the windward leading edge of the LMC disc. 

\citet{vdmarel2001} showed that the LMC has an intrinsic  ecccentricity in its outer part, with an
elongation in the general direction of the Galactic centre  induced by the tidal 
force of the MW. Note that such an outer disc is off-centred from both the LMC bar and the kinematical 
centre \citep[see figure 1 in][]{beslaetal2016}.  The different ratios we found between 
radii estimated from DES data and in this work also support the notion that old populations are 
more elongated than young populations \citep[see also][]{jacyszynetal2017}.  Because old
populations prevail in the outermost northern LMC fields, we infer that
the MW tidal forces that caused such an elongation should not have acted recently, as for example, 
during the claimed first passage of the LMC to the MW \citep{beslaetal07,kallivayaliletal13}.
Note that such an elongation is also seen in the deprojected spatial distribution of clusters 
with ages between 1-3 Gyr (ratio between northern and western radii $\sim$ 2.2) \citep{p14b}. 

On the other hand, the similar radii found in this work for young and old populations
along the western LMC minor axis  confirm previous 
suggestions that ram pressure stripping could cause that the gaseous LMC volume moved windwards 
due to the motion of the LMC in the MW halo. This outcome also
explains the recent (age $\la$ 50 Myr) important star cluster formation activity that took place
in the outermost western LMC regions \citep{petal2017}.

\section{Conclusions}

We used Washington $MT_1$ photometry of 15 LMC fields (36$\arcmin$$\times$36$\arcmin$ wide each)
distributed along the western half part of the minor axis of the LMC, and two additional reference
MW fields, 
to trace the stellar density profile of LMC young and old populations. The whole sample of selected 
fields expands a baseline in deprojected distances from the LMC bar centre up to $\sim$ 31.6 kpc
away. 

In order to perform the count of stars we built CMDs and defined two regions: a strip
along the young MS (age $<$ 1 Gyr) and a RC box for intermediate-age and old populations
($\ge$ 1 Gyr). We took into account incompleteness  effects in our photometry, reddening effects and 
the contamination from the MW foreground field.  From the analysis of the resultant
profiles, we draw the following conclusions:

$\bullet$ Both young
and old populations reach similar LMC boundaries, with a mean value of 8.9 $\pm$ 0.4 kpc.

 $\bullet$ This LMC limit is significantly smaller than those found
in previous studies of the  extension of the LMC, which range between 15.0 and 18.5 kpc. 

$\bullet$ Old populations are nearly three times more dense that young populations
at that LMC limit. 

$\bullet$ When comparing our values with those derived towards the north using the DES database, we
found that, after correcting for projection effects, the old populations are significantly
more elongated than the young ones. \citet{vdmarel2001} had suggested that outer LMC regions
are more elongated than the inner ones in the general direction of the Galactic centre
because of the tidal forces of the MW. Here we confirm that pattern, that additional shows
to be 
age-dependent: old populations prevail in the most elongated stellar distributions.
The similar radii found for young and old populations along the western half 
part of the LMC minor axis also agree well with this picture.

$\bullet$ The existence of young populations in the western outermost regions of the LMC is, in addition,
an observational evidence of more recent ram pressure stripping due to the motion of the LMC in
the MW halo. Because of that, the gaseous LMC disc shows a truncated density profile along the 
windward leading edge (east-northeast direction)  and young field star and cluster populations
in the western side of the LMC.

\section*{Acknowledgements}
We thank the referee for his/her thorough reading of the manuscript and
timely suggestions to improve it. 




\bibliographystyle{mnras}

\input{paper.bbl}







\bsp	
\label{lastpage}
\end{document}